\documentclass[reprint, superscriptaddress,
 amsmath,amssymb,
 aps,pra]{revtex4-2}

\usepackage{graphicx}
\usepackage[english]{babel}
\usepackage[utf8x]{inputenc}
\usepackage{dcolumn}
\usepackage[colorinlistoftodos]{todonotes}
\usepackage{xcolor}
\usepackage{mathtools}
\usepackage{bm}
\usepackage{hyperref}
\usepackage{psfrag}
\usepackage{indentfirst} 
\usepackage{url}
\usepackage{enumitem}
\usepackage{tensor}
\usepackage{braket}
\usepackage{bbm}
\usepackage{dsfont}
\usepackage{amsfonts,latexsym,cancel}
\usepackage{rawfonts}
\usepackage{float}
\usepackage{pictexwd}
\usepackage{wrapfig}
\usepackage{subfigure} 
\usepackage{soul}

\usetikzlibrary{arrows}

\begin{document}

\preprint{}

\title{Adiabatic lapse rate of real gases}

\author{Bogar Díaz}
\email{bodiazj@math.uc3m.es}
\affiliation{Departamento de Matem\'aticas, Universidad Carlos III de Madrid. Avenida  de la Universidad 30, 28911 Legan\'es, Spain}
\affiliation{Grupo de Teor\'ias de Campos y F\'isica Estad\'istica. Instituto Gregorio Mill\'an (UC3M), Unidad Asociada al Instituto de Estructura de la Materia, CSIC, Serrano 123, 28006 Madrid, Spain}
\author{Miguel \'Angel Garc\'ia-Ariza}
\email{mgarciaariza@astate.edu}
\affiliation{Arkansas State University Campus Quer\'etaro. Carretera Estatal 100 km 17.5, 76270 Colon, Qro., Mexico} 
\author{J. E. Ram\'irez}
\email{jhony.ramirezcancino@viep.com.mx}
\affiliation{Centro de Agroecolog\'ia, Instituto de Ciencias, Benem\'erita Universidad Aut\'onoma de Puebla, Apartado Postal 165, 72000 Puebla, Pue., Mexico} 
\affiliation{Consejo de Ciencia y Tecnolog\'ia del Estado de Puebla, Privada B poniente de la 16 de Septiembre 4511, 72534, Puebla, Pue., Mexico}

\begin{abstract}
We derive a formula for the dry adiabatic lapse rate of atmospheres composed of real gases. We restrict our study to those described by a family of two-parameter cubic equations of state and the recent Guevara non-cubic equation. Since our formula depends on the adiabatic curves, we compute them all at once, considering molecules that can move, rotate, and vibrate, for any equation of state. To illustrate our results, we estimate the lapse rate of the troposphere of Titan, obtaining a better approximation to the observed data in some instances, when compared to the estimation provided by the virial expansion up to third order. 
\end{abstract}

\maketitle


\section{Introduction}
The \textit{adiabatic lapse rate} of an atmosphere, denoted henceforth by $\Gamma$, is defined to be the rate of change of temperature with respect to height, assuming that parcels of atmosphere at constant height are in equilibrium and that there is no heat exchange between any two of them. This number has been computed directly from observations in many objects within the solar system~\cite{Catling2015}. Nonetheless, its theoretical reproduction is rather involved. This is in part because real atmospheres are composed of gas mixtures containing traces of vapor or pollutants, and also because of the many thermodynamic processes within them.

In this work, we attempt to improve previous theoretical approximations to the adiabatic lapse rate of dry  atmospheres (DALR) by incorporating into its computation \textit{complete} equations of state (EOS), rather than truncated virial expansions \cite{JHON, LR2}. This is applied to the atmosphere of Titan, whose surface atmospheric conditions are close to the boiling point of $\text{N}_2$ ($P\approx 101\,\text{kPa}$, $T\approx77\,\text{K}$).

The first approach to the theoretical computation of the adiabatic lapse rate of atmospheres considers the ideal gas model \cite{kittel198}. As expected, this yields results far from observations \cite{Catling2015}, because this EOS overlooks molecular vibrations and interactions within the gas, among other reasons. Some authors have tried to overcome these downturns by considering ideal gases, but experimental data for heat capacities at constant pressure \cite{Catling2015, Vallero2014}. To some extent, this incorporates information on interactions and molecular vibrations. However, from the conceptual point of view this is an incorrect procedure because the used lapse rate formula strongly depends on the adiabatic curves \cite{JHON, LR2}. Another method to derive a DALR for real gases consists of considering the experimental values of the compressibility factor \cite{Staley1970}. Although this scheme yields a good approximation to the experimental value for Venus, it is limited by the availability of experimental data in the range of atmospheric conditions for other astronomic objects. 

The shortcomings of resorting to experimental values, as described above, may be overcome by using a \textit{truncated} virial expansion and modifying the adiabatic curves to incorporate both intermolecular interactions and molecular vibrations \cite{JHON, LR2}. This method improves the value of $\Gamma$ significantly in extreme atmospheric conditions and allows one to identify the origin of the correction, namely, molecular vibrations (like in Venus's atmosphere), molecular interactions (like in Titan's atmosphere), or both \cite{LR2}.

This work extends the procedures described above by considering complete EOS, in order to further understand the relevance of each ingredient that makes up the DALR of an atmosphere. In particular, we are interested in the role of interactions beyond the two- and three-particle ones. These are taken into account by means of the compressibility factor, denoted by $Z$.

We work with two-parameter cubic EOS and an extension of this family, known as the Guevara non-cubic EOS. Our choice of EOS relies on the fact that the family we pick is relatively simple and, at the same time, correctly reproduces certain experimental data on the behavior of real fluids \cite{LOPEZECHEVERRY201739}. An advantage of our approach is, however, that the formula we obtain for the DALR can be used with any EOS, regardless of the motivation underlying its choice.

We have organized this paper as follows. In Sec. \ref{sec2}, we review the formula for the lapse rate in terms of $Z$, as obtained in Ref. \cite{LR2}, and apply it to the EOS of interest. The resulting expressions depend on the adiabatic curves of each model, which are computed in Sec. \ref{adcurves}. In Sec. \ref{results}, we plug into our formulas the information to calculate the lapse rate of Titan and compare the theoretical $\Gamma$ of each EOS with observed data and  with the prediction of the virial expansion up to the third order. We wrap up this work with our concluding remarks in Sec. \ref{sec:conclu}.

\section{Dry adiabatic Lapse rate}\label{sec2}
In this section, we discuss the derivation of the DALR formula presented in Ref. \cite{LR2} and its application to a family of cubic EOS and the non-cubic Guevara EOS.

As we said before, the lapse rate of the atmosphere is defined as 
\begin{equation}
    \Gamma =\frac{\mathrm{d} T}{ \mathrm{d} z}\,,
\end{equation}
where $T$ is the temperature and $z$ is the height ranging within the troposphere of the planet.
To obtain the formula for $\Gamma$ as in Ref. \cite{LR2}, it is necessary to relate the derivatives of the compressibility factor $Z=PV/(RT )$ on adiabatic curves to the hydrostatic equation $\mathrm{d} P/ \mathrm{d} z=-M_\text{mol}g/V$.
Here, $P$ is the pressure at height $z$, $V$ is the volume corresponding to given values of temperature and pressure, $M_\text{mol}$ is the molar mass of the atmosphere, $g$ is the magnitude of the acceleration near to the planet surface (its variation across the troposphere is usually less than 1\%), and $R$=8.31 J mol$^{-1}$ K$^{-1}$ is the gas constant. We assume atmospheres to be homogeneous monocomponent simple fluids (consisting of the most abundant gas), whence the compressibility factor depends on volume and temperature only. On adiabatic curves, $Z$ can be expressed solely in terms of the temperature. As usual, we consider that neither the temperature nor the lapse rate itself depends on latitude.

From the definition of the compressibility factor, we obtain that, on adiabatic curves, 
\begin{equation}\label{eq:adiabaticP}
\mathrm{d} P=\frac{R}{V}\left( Z+T\frac{\partial Z}{\partial T}-Z T\frac{V'}{V}+T V'\frac{\partial Z}{\partial V}  \right) \mathrm{d} T\,.
\end{equation}

In what follows, we denote the adiabatic volume and the adiabatic compressibility factor as $V_\text{ad}$ and $Z_\text{ad}, $ respectively. 
In addition, primed variables denote their derivative with respect to temperature, for example, $V'=d V/d T$.
Using Eq. \eqref{eq:adiabaticP} and the hydrostatic equation, we have that the adiabatic lapse rate can be computed as
\begin{align}
    \Gamma=&\Gamma^\text{IG}\frac{C_P^\text{IG}}{R}\left( Z_\text{ad}+T \left. \frac{\partial Z}{\partial T}\right|_{V_\text{ad}}\right.\nonumber\\
    &\left. \qquad \qquad -Z_\text{ad}T\frac{V_\text{ad}'}{V_\text{ad}} +TV_\text{ad}' \left. \frac{\partial Z}{\partial V} \right|_{V_\text{ad}} \right)^{-1}\,,
\label{eq:LRZad}
\end{align}
where $\Gamma^\text{IG}=M_\text{mol}g/C_P^\text{IG}$ is the adiabatic lapse rate for ideal gases and $C_P^\text{IG}$ is the heat capacity at constant pressure, expressed as $(5+f_r)R/2$, with $f_r$ being the rotational degrees of freedom of the gas molecules, which takes the value 0, 2, and 3 for monoatomic, diatomic or linear, and polyatomic molecules, respectively. We must mention that an alternative to Eq. \eqref{eq:LRZad} is  presented in Ref. \cite{Lindal1983}. In the latter, the heat capacity at constant pressure obtained from the EOS is an input to obtain $\Gamma$, whereas in Eq. \eqref{eq:LRZad}, only the $C_p$ of the ideal gas is required. This is an advantage, due to its simple form. Besides, $Z$ is more common than a the correcting factor of temperature used in Ref. \cite{Lindal1983}.

As we mentioned before, for Eq. \eqref{eq:LRZad}, the compressibility factor can be found from tabulated experimental data. This procedure yields a lapse rate that incorporates all the information of the molecules within the gas, including interactions and vibrations. Nevertheless, a shortcoming of this approach is the availability of data for the atmospheric conditions of the astronomical objects under study.

A theoretical alternative is considering a particular model for the EOS, namely, ideal gas ($Z=1$), van der Waals, Redlich-Kwong, etc. This is the path we follow in the present paper. As we stated in the Introduction, we are interested in the well-known family of two-parameter cubic EOS.
These equations of state are relevant because they incorporate information about molecular interactions and their ranges. Moreover, they predict phase transitions. This last feature could be of significant importance in the computation of lapse rates for astronomical objects having atmospheres under conditions near to critical points, as is the case of Titan, whose surface atmospheric conditions are close to the Nitrogen boiling point.

The general two-parameters cubic EOS  is \cite{wenzel}
\begin{equation}
    P=\frac{R T}{V-b} -\frac{a(T)}{V^2+u b V+w b^2}\,.
    \label{eq:ceos}
\end{equation}
Several cubic equations of state can be reproduced by setting the adequate value of $u$ and $w$, and the functional form for $a(T)$.
In this work, we are interested in the van der Waals, Redlich-Kwong, and Peng-Robinson models. We also include other interesting cubic EOS, developed by Guevara in Ref.~\cite{cg}, wherein the author constructs a square-well-type fluid model whose reduced EOS fulfills the corresponding-states principle.
Remarkably, Guevara's cubic EOS reproduces very well the coexistence diagram for simple fluids, even water. We use the fact that all these equations of state can be written as
\begin{equation}
    P=\frac{RT}{V-b}-\frac{a(T)}{(V-c)(V-d)}\,.
    \label{eq:fact}
\end{equation}

In Table \ref{tab:uw} we show the information concerning the mentioned equations of state.

\begin{table*}
\begin{center}
\begin{tabular}{c c c c c c}
\hline
EOS &$u$ & $w$ & $c$ & $d$ & $a(T)$\\
\hline
van der Waals & 0 & 0 & 0 & 0 & $a_c$\\
Redlich-Kwong & 1 & 0 & -$b$ & 0 & $a_cT^{-1/2}$ \\
Peng-Robinson & 2 & -1 & $(-1+\sqrt{2})b$ & $(-1-\sqrt{2})b$ & $a_c[1+\kappa(1-T_r^{1/2})]^2$\\
Guevara &  $-(a_1+a_2)/b$ &   $a_1 a_2/b^2$ &  $a_1$ & $a_2$ & $RTb(\lambda^3-1)( \exp{[\varepsilon/(k_B T)]}-1)$ \\
\hline
\end{tabular} 
\end{center}
\caption{Parameters of the EOS. Some of these admit a physical interpretation: $b$ corresponds to the excluded volume, $a_c$ is the magnitude of the intermolecular repulsive force, $T_r$ denotes the reduced temperature, $\kappa$ is related to the acentric factor, and $\lambda$ and $\varepsilon$ are the range and the amplitude of the attractive interaction of the square-well potential.}\label{tab:uw}
\end{table*}


Recently, Guevara also developed a non-cubic extension of his cubic EOS (denoted here by $P_0$) given by
\begin{equation}
    P_\text{G}=P_0-\frac{f T}{(V-e)^\nu}\,,
    \label{eq:ncG}
\end{equation}
where $e,f$, and $\nu$ are additional parameters that are adjusted according to the substance to be studied \cite{ncG}. This non-cubic EOS predicts phase diagrams for simple fluids that are in agreement with the experimental data, which renders it particularly suited for the lapse rate computation in situations where the tropospheric conditions are close to boiling or sublimation points, like in Titan's atmosphere, as we show below.

As we mentioned before, we aim at studying the effect that complete equations of state have on the computation of the lapse rate, compared to the results obtained with a truncated virial expansion. To this end, we recall that the virial expansion of an equation of state is
\begin{align}
P_\text{v}= \frac{RT}{V}\left( 1+\sum_{k=1}\frac{B_{k+1}}{V^k} \right)\,, \label{eq:VE}
\end{align}
where $B_{k+1}$ are the so-called virial coefficients. These functions are generically sums of the cluster integrals involving the $(k+1)$-particle interactions \cite{mayerbook}. By construction, they \textit{only} depend on temperature \cite{mcquarriethermo} in states sufficiently far from the boiling point \cite{ushcats4}. In Sec.~\ref{results}, we  also compare results for the equations of state discussed above with its corresponding virial expansion up to the third coefficient.
In particular, the virial coefficients of Eq. \eqref{eq:fact} are
\begin{equation}
    B_k(T)=b^{k-1} - \frac{a(T)}{R T} (c + d)^{k-2}\,,
\label{eq:virials}
\end{equation}
for $k> 2$. In the case of the van der Waals EOS ($k=2$), we have $B_2(T)=b - a_c/(R T) $. The first 12 virial coefficients of the non-cubic and cubic Guevara EOS coincide. Notice that the truncated virial expansions also depend on the same parameters ($a_c,b,c$,\textit{ etc.}) of the original EOS, but they \textit{only} contain information on $k$-particle interactions. This is an important physical difference between  truncated virial expansions and complete EOS.

The compressibility factor for the cubic equations of state \eqref{eq:fact}, Guevara's non cubic equation \eqref{eq:ncG}, and the virial expansion \eqref{eq:VE} are 
\begin{subequations}
\begin{align}
    Z&=\frac{V}{V-b} -\frac{a(T)V}{RT (V-c)(V-d)}\,,\\
    Z_G&=  Z_0-\frac{fV}{R(V-e)^\nu} \,,\\
    Z_\text{v}&=1+\sum_{k=1}\frac{B_{k+1}}{V^k}\,,
\end{align}
\end{subequations}
respectively, where $Z_0$ represents the compressibility factor of Guevara's cubic EOS. Then, the lapse rates in these cases are
\begin{widetext}
\begin{subequations}\label{eq:LR}
\begin{align}
    \Gamma&=\Gamma^\text{IG} \frac{C^{IG}_P}{R} \left( \frac{V_\text{ad}}{V_\text{ad}-b} -\frac{a'(T)V_\text{ad}}{R(V_\text{ad}-c)(V_\text{ad}-d)} + V_\text{ad}V_\text{ad}' \left( \frac{a(T)(2V_\text{ad}-c-d)}{R(V_\text{ad}-c)^2(V_\text{ad}-d)^2}-\frac{T}{(V_\text{ad}-b)^2}  \right) \right)^{-1} \,, \label{eq:LRc}\\
    \Gamma_\text{G}&=  \Gamma_0\left( 1+\frac{\Gamma_0 R}{\Gamma^\text{IG}C_P^\text{IG}} \left( V_\text{ad}V'_\text{ad}\frac{Tf\nu}{R(V_\text{ad}-e)^{\nu+1}} -\frac{fV_\text{ad}}{R(V_\text{ad}-e)^\nu}\right)  \right)^{-1}\,,\label{eq:LRG}\\
\Gamma_\text{v}&=\Gamma^\text{IG} \frac{C^{IG}_P}{R} \left( 1+\sum_{k =1}\frac{\mathcal{B}_{k+1}}{V_\text{ad}^k} -T\frac{V_\text{ad}'}{V_\text{ad}}\sum_{k=1}\frac{(k+1)B_{k+1}}{V_\text{ad}^k}\right)^{-1} \,, \label{eq:LRv}
\end{align}
\end{subequations}
with $\mathcal{B}_{k+1}=TB'_{k+1}+B_{k+1}$ and $\Gamma_0$ denoting the lapse rate corresponding to the Guevara cubic EOS. Notice that $V_\text{ad}$ is the adiabatic volume for the corresponding EOS, which we derive in the next section.
\end{widetext}

\section{Adiabatic curves}\label{adcurves}

As we have stated, to calculate the DALR we need the adiabatic curves corresponding to the EOS under consideration. In this section, we show in a general framework, that it is possible to obtain, at least in quadratures, the equation of the adiabatic curves of \textit{any} equation of state. 

Given an equation of state $P=P(V,T)$, the corresponding Helmholtz free energy $A$ is calculated as \cite{zemansky1997heat,Matsumoto2005}
\begin{equation}\label{eq:Hfe}
    A=-\int P \mathrm{d} V+\phi(T)\,,
\end{equation}
with \footnote{In the calculation of $\phi(T)$ there is an arbitrary function $f(T)$ that comes from the derivation of the contribution of the molecular interactions to the canonical partition function. However, it is fixed to zero in order to reproduce the adiabatic curves for the virial expansion reported in Ref.~\cite{LR2}.}
\begin{equation}
    \phi(T)=k_BT\ln N!-Nk_BT\ln \left( \frac{q_\text{tras}}{V} q_\text{rot} q_\text{vib}  \right)\,, \label{eq:phi}
\end{equation}
where
\begin{subequations}\allowdisplaybreaks
\begin{align}
    \frac{q_\text{trans}}{V}&= \left( \frac{2\pi M k_B T}{h^2} \right)^{3/2}\,, \label{eq:pf2}\\
    q_\text{rot}&=  \frac{T^{f_{r}/2}}{\theta_\text{rot}}\,,  \label{eq:pf3}\\
    q_\text{vib}&=  \prod_{j=1}^m \frac{\exp(-h\nu_j/(2k_BT))}{1-\exp(-h\nu_j/(k_BT))} \,,  \label{eq:pf4}
\end{align}\label{eq:fpmi}
\end{subequations}
represent the partition functions corresponding to translations, rotations, and vibrations, respectively \cite{mcquarrie,Mayer1958,mayerVirial}. In the equations above, $M$ is the molecular mass, $h$ is the Planck constant, $\theta_\text{rot}$  and $\nu_1,\ldots,\nu_m$ are the characteristic rotational temperatures and the natural vibrational frequencies of the gas molecules, respectively. As is well known, for molecules with $n$ atoms, $m=3n-1$ (linear molecules) or $m=3n-6$ (nonlinear molecules) \cite{mcquarrie}. Notice that $q_\text{vib}$ is a function that  depends on $T$ and on the vibrational properties of the gas under consideration. 


On the other hand, from statistical mechanics we have that the Helmholtz energy is calculated by $A=-k_BT\ln Q$, where $Q$ is the canonical partition function of the system. Comparing this with Eq. \eqref{eq:Hfe}, we get
\begin{equation}
    \ln Q=\frac{1}{k_B T}\int P \mathrm{d} V - \tilde{\phi}(T)\,,
\end{equation}
with $\tilde{\phi}(T)=\phi(T)/(k_B T)$. With this information we can calculate the internal energy $U$ as
\begin{align}
U&=k_\text{B}T^2 \left( \frac{\partial \ln Q}{ \partial T}\right)_V\nonumber\\
&= \int \left( T\frac{\partial  P}{\partial T}-P  \right) \mathrm{d} V-k_BT^2\tilde{\phi}'(T)\,. 
\end{align}
Integrating the adiabatic curves, $\mathrm{d} U + P \mathrm{d} V = 0$, of the system in the $T-V$ representation yields 
\begin{equation} \label{eq:pac}
\int \frac{\partial P}{\partial T}  \mathrm{d} V-\int \frac{1}{T}\frac{\mathrm{d} }{\mathrm{d} T} \left( k_B T^2 \tilde{\phi}'(T)  \right) \mathrm{d} T =\text{const.}
\end{equation}
Making use of Eq. \eqref{eq:phi}, Eq. \eqref{eq:pac} for 1 mole of gas is
\begin{equation}
\int \frac{\partial P}{\partial T} \mathrm{d} V+ C^{\text{IG}}_V \ln T+\frac{R T q'_\text{vib}}{q_\text{vib}}+R \ln (q_\text{vib})  =\epsilon_0\,, \label{eq:adc}
\end{equation}
where we have identified the heat capacity of ideal gases at constant volume $C^{\text{IG}}_V=(3+f_r)R/2 $, and all constants are grouped in $\epsilon_0$. We remark that the heat capacity $C_V$ corresponding to the EOS does not necessarily coincide with that of the ideal gas. It is, rather, obtained by differentiating $U$ with respect to $T$ at constant volume, which is in general not constant.

Equation \eqref{eq:adc} is the general expression in quadratures of the adiabatic curves of 1 mole of gas. This expression coincides with others that have been reported before \cite{Tjiang_2006} up to an arbitrary function of temperature. The latter has been fixed herein by considering the microscopic degrees of freedom of the molecules.

Using Eq. \eqref{eq:adc}, we have that the adiabatic curves for the cubic \eqref{eq:fact}, viral \eqref{eq:VE}, and Guevara non-cubic  \eqref{eq:ncG} EOS can be written respectively as
\begin{subequations}
\begin{align}
\left(\frac{V-d}{V-c}\right)^{a'(T) R/\left[ (c-d) C_V^{IG} \right]}
     W(V,T)&=\varepsilon_0\,,\\
    T\left[Vq_\text{vib}
    \exp\left( \frac{ T q'_\text{vib}}{q_\text{vib}} - \sum_{k=1} \frac{\mathcal{B}_{k+1}}{kV^k} \right) \right]^{R/C_V^{IG}}&=\varepsilon_0 \,,\label{eq:e0}\\
    \mathcal{V}_0(V,T)  \exp\left( \frac{f}{(\nu-1)(V-e)^{(\nu-1)}}\right) &=\varepsilon_0\,,
\end{align}
\end{subequations}
where
\begin{equation}
W(V,T)=   T\left[(V-b)q_\text{vib}
    \exp\left( \frac{ T q'_\text{vib}}{q_\text{vib}} \right) \right]^{R/C_V^{IG}}
\end{equation}
for all but van der Waals' EOS, for which $W(V,T)=\varepsilon_0$, and $\mathcal{V}_0(V,T)$ is the functional form of the adiabatic curves for the Guevara cubic EOS. 
Equation \eqref{eq:e0} coincides with the result reported in Ref. \cite{LR2}. We point out that, in the equations above, we incur in an abuse of notation, since the parameter $b$ and the constant $\varepsilon_0$ depend on the equation of state.

To close this section, we emphasize that translations, vibrations, and rotations of molecules contribute to the internal energy and the heat capacity of the gas, and hence shape the adiabatic curves accordingly. This can be seen explicitly in Eq. \eqref{eq:adc}. However, the relevance of each kind of motion depends on the given atmospheric conditions. In particular, the contribution of vibrations becomes relevant when the temperature is comparable to one of the vibrational temperatures of the gas. For instance, in Venus' atmosphere, vibrations are the principal correcting factor for the lapse rate, even when considering truncated virial expansions, as was shown in Ref. \cite{LR2}. This is true because the surface temperature of Venus (737 K) is close to one of the vibrational temperatures of CO$_2$ ($\sim$960 K). In contrast, the vibrational temperature of N$_2$ is 3349 K, which is 2 orders of magnitude above Titan's surface temperature (94 K). Under such circumstances, molecular vibrations are negligible in the computation of the lapse rate of Titan's atmosphere \cite{LR2}. This is not the case for molecular interactions, as we illustrate below.

\section{Estimations for Titan}\label{results}

As was shown before \cite{LR2}, the contribution to the DALR from molecular interactions becomes relevant in atmospheric conditions close to phase transitions. For this reason, we have chosen the atmosphere of Titan, which is close to the boiling point, to illustrate how the estimation of the DALR obtained from a complete EOS compares to that provided by its corresponding third-order virial expansion, in reference to the observed value. To this end, we discuss the procedure to compute the parameters of the EOS, the adiabatic curves, and the lapse rate of Titan's atmosphere, considered to be composed exclusively of N$_2$ (the actual concentration thereof is 92\%  \cite{Catling2015}). It is important to remark that contributions from the consideration of mixtures are expected to be negligible in this case. For instance, from the point of view of virial expansions, the second virial coefficient of a binary mixture may be written as $B=y_1^2B_1+y_1y_2B_{12}+y_2^2 B_2$, where the $y_i$'s and the $B_i$'s are molar fractions and the second virial coefficients of the $i$-th component, respectively, and $B_{12}$ is an interaction coefficient \cite{Dy2003mix}. Small molar fractions amount to even smaller second and third terms in this expression. The same holds for higher-order virial coefficients, which involve higher-order dependencies on molar fractions. The latter is not exclusive to Titan, but is the case of other celestial bodies like Venus (96.5\% CO$_2$) and Mars (96\% CO$_2$), where a monocomponent approach is not only reasonable but also sufficient. On the other hand, however, mixtures may be relevant in objects like Saturn (88\% H$_\text{2}$ vs. 12\% He).

\subsection{Parameters of the EOS}\label{sec:parameters}
We obtain the values of some of the parameters appearing in the EOS \eqref{eq:fact} and \eqref{eq:ncG}. We use the equations established for the temperature, pressure, and volume at the critical point for N$_2$, considering 1 mol of gas. Correspondingly, we take $T_c=126.19\,\text{K}$, $P_c=3.3978\,\text{MPa}$, and $1/v_c=11.177\,\text{mol/dm}^3$ \cite{datos}. We report in Table \ref{tab:parameters} all the EOS parameters obtained with these data.

\begin{table*}
\begin{center}
\begin{tabular}{c c c}
\hline
EOS & Parameters & Values \\
\hline
van der Waals & $a_c=27bRT_c/8$ & 0.104759 m$^6$Pa \\
 & $b=V_c/3$ & 0.0000296 m$^3$\\
Redlich-Kwong & $a_c=0.42748 R^2 T_c^{2.5}/P_c$ & 1.5541m$^6$PaK$^{0.5}$ \\
 & $b=0.08664 R T_c/P_c$ & 0.0000267m$^3$\\
Peng-Robinson & $a_c=0.45724R^2 T_c^2/P_c$  & 0.147978m$^6$Pa \\
 & $b=0.0778R T_c/P_c$ & 0.000024m$^3$\\
 Guevara & $b=(Z_c-1+\alpha_c)V_c/Z_c$ & 0.000033m$^3$ \\
 & $a_1=(Z_c-0.5\alpha_c+\alpha_c\sqrt{\alpha_c-0.75}) V_c/Z_c$ & 0.0000286m$^3$\\
 & $a_2=(Z_c-0.5\alpha_c-\alpha_c\sqrt{\alpha_c-0.75})V_c/Z_c$ & -0.0001031m$^3$\\
 & $Z_c=P_cV_c/RT_c$ & 0.28837 \\
 & $\alpha_c^3=a(T_c)P_c/R^2T_c^2$ & 0.7863 \\
\hline
\end{tabular} 
\end{center}
\caption{Values of the EOS parameters calculated from the critical point.}\label{tab:parameters}
\end{table*}

The value of $\kappa$ in the Peng-Robinson EOS is 0.4324, using that $\kappa=0.37464+1.54226\omega-0.26992\omega^2$ and that $\omega=0.03772$ for $\text{N}_2$\cite{PR, ncG}. Guevara's cubic EOS requires the computation of the parameters of $a(T)$, which coincide by construction with those of the square-well potential, i.e., $\lambda=2.71622$ and $\varepsilon/k_B=30.0625$ K$^{-1}$ \cite{cg}. For the non-cubic term in Eq. \eqref{eq:ncG}, we take $e=2.98623 \times 10^{-5}$ m$^3$, $f= 5.95905 \times 10^{-67}$ Pa(m$^3$/mol)$^{12}$K$^{-1}$, and $\nu=12$, as reported in Ref. \cite{ncG}.

\subsection{Computation of the adiabatic curves}\label{sec:adc}

To calculate the values of $\varepsilon_0$ in each case, and thus obtain the adiabatic curves, we use the surface atmospheric conditions of Titan, $T_S=94$ K and $P_S=146.7$ kPa. These are reported in Table \ref{tab:e0}. We also consider that the wave number associated with the vibrational frequencies of $\text{N}_2$ is 2328.72 cm$^{-1}$ \cite{dataN2}.

The values of $V_\text{S}$ are determined by finding physically admissible solutions (real and greater than the excluded volume $b$) for $P_\text{S}=P(V,T_\text{S})$ considering 1 mol of gas.

\begin{table}[!ht]
\begin{center}
\begin{tabular}{c c c}
\hline
EOS &  $\varepsilon_0$ \\
\hline
van der Waals &  11.4591 \\
Virial-van der Waals & 11.4605\\
Redlich-Kwong &  11.3029 \\
Virial-Redlich-Kwong &  11.3029\\
Peng-Robinson &  11.3209\\
Virial-Peng-Robinson & 11.3209\\
Guevara & 11.1833\\
Virial-Guevara & 11.3305\\
nc-Guevara & 11.0777 \\
\hline
\end{tabular} 
\end{center}
\caption{Values of $\varepsilon_0$.  \label{tab:e0}}
\end{table}

Notice that the only model in which $V=V(T)$ can be written in a closed form is the van der Waals EOS. For the remaining EOS, we use Newton's method with $\Delta T=0.01$. The corresponding derivative is computed by means of the five-point stencil method, with spacing $h=0.001$. The resulting numerical data are implemented into the lapse rate by fitting $V(T)$ and $V'(T)$ to cubic splines. Thereby, we obtain continuous expressions for the DALR.

\subsection{Lapse rate}

\begin{figure}[ht]
\centering
\includegraphics[scale=1]{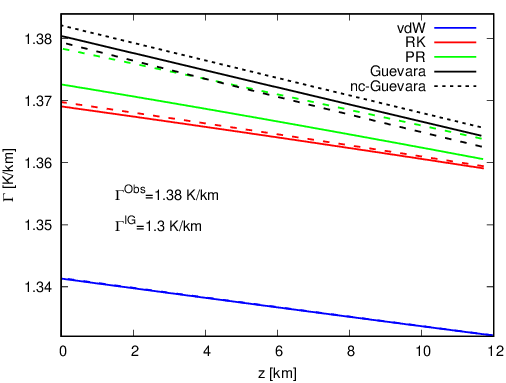}
\caption{DALR corresponding to different EOS (solid line) and their corresponding virial expansions up to third order (dashed line) : van der Waals (blue line), Redlich-Kwong (red line), Peng-Robinson (green line), Guevara (black line), and Guevara non-cubic (dotted line).}
\label{fig:lr-all}
\end{figure}

With the results of the previous sections, we may now obtain the DALR of Titan as a function of height. Indeed, substituting in Eq. \eqref{eq:LR} the information of the adiabatic curves obtained in Sec. \ref{sec:adc}, we find $\Gamma$ as a function of $T$. By solving $dT/dz=\Gamma$ for $T$ using the fourth-order Runge-Kutta method, we may write $T=T(z)$ and thus express the DALR in terms of the height. We present our results in Fig. \ref{fig:lr-all}, where we use $g=1.35\text{m}/\text{s}^2$.

\begin{figure*}[ht]
\centering
\includegraphics[scale=1]{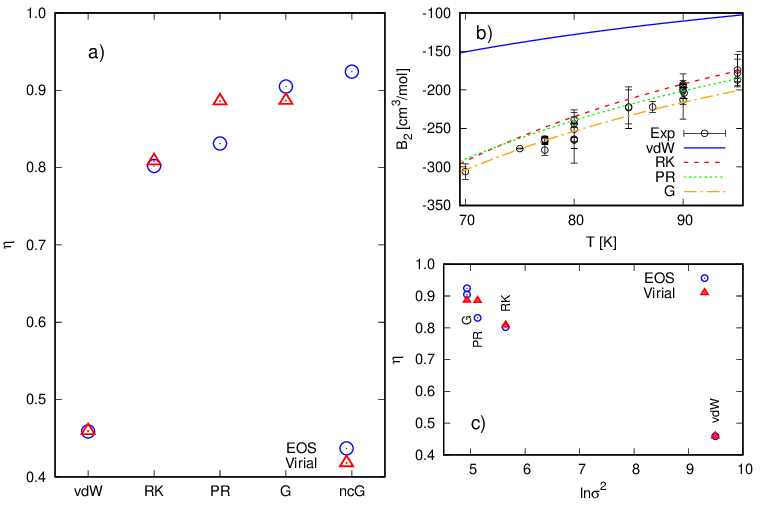}
\caption{a) The values of $\eta$ corresponding to different EOS (circles) and and their corresponding virial expansions up to third order (triangles). b) Theoretical (lines) and experimental (circles) $B_2$. c) $\eta$ vs. $\sigma^2$ for complete EOS (circles) and virial expansions (triangles).} 
\label{fig:eta}
\end{figure*}

In order to measure the accuracy of the theoretical value of $\Gamma$ obtained from each EOS, we compare the observed lapse rate $\Gamma^{\text{Obs}}$ to the average rate of change of $T(z)$ obtained with $\Gamma$, which we call $\overline{\Gamma}$. We do this because $\Gamma$ is not constant, whereas $\Gamma^{\text{Obs}}$ is. Then, we define
\begin{equation}\label{eta}
\eta = \frac{\Gamma^\text{IG}-\overline{\Gamma}}{\Gamma^{\text{IG}}-\Gamma^{\text{Obs}}}\,,
\end{equation}
which takes values around 0 when $\overline{\Gamma}$ is close to $\Gamma^\text{IG}$ and values close to 1 when $\overline{\Gamma}$ is a good approximation to $\Gamma^\text{Obs}$. In Fig.~\ref{fig:eta} a) we show the values of $\eta$ for the EOS under consideration. We also compare them to the values of $\eta$ of their corresponding virial expansions.

As can be seen, some models yield a better approximation to $\Gamma^\text{Obs}$ than others. We conjecture that the accuracy of $\Gamma$ is related to the precision with which an EOS reproduces the observed second virial coefficient. To substantiate this, we first determine how accurately a model reproduces $B_2$ [see Fig.~\ref{fig:eta} b)], which is done by using the mean square distance $\sigma^2$ between the theoretical $B_2$ [see Eq.~\eqref{eq:virials}] and the experimental data in Ref. \cite{dymond2002virial} within the temperature range where the DALR was calculated. As depicted in  Fig.~\ref{fig:eta} c), $\eta$ improves as $\sigma^2$ decreases, suggesting a possible criterion for the choice of an EOS that accurately reproduces the DALR of Titan.

\section{Conclusions}\label{sec:conclu}

The formula for the DALR of atmospheres presented in this work represents an improvement to previous considerations in the literature, where intermolecular interactions are taken into account through a truncated virial expansion. In contrast, we have considered complete EOS that describe real gases, namely, a family of two-parameter cubic equations of state and the Guevara non-cubic equation. It is important to remark that the DALR depends on the adiabatic curves of the corresponding EOS, for which we have obtained a general expression. In this step, we have considered moving molecules that rotate and vibrate. In those cases where we were not able to solve explicitly for the adiabatic volume, we have implemented numerical methods. We remark that an advantage of our approach is that Eq. \eqref{eq:LRZad} can be used with any EOS, regardless of what motivates its choice.

The expression we obtain for the lapse rate can be applied to any atmosphere if we have the required input. For example, it can be used to estimate the DALR for exoplanets (see Ref. \cite{LR2}). We chose Titan's because its surface atmospheric conditions are close to those allowing for phase transitions, where contributions to the lapse rate from intermolecular interactions are known to be relevant. Notice that the role of molecular vibrations in Titan's atmosphere is not relevant. This is because the surface temperature of its atmosphere (94 K) is far from the vibrational temperature of $N_2$ ($\approx$ 3349 K) \footnote{For instance, the contributions of the vibrations to the heat capacity [third and fourth term in \eqref{eq:adc}] range across the troposphere from $4.12\times 10^{-13}$ J/K to $4.48\times 10^{-17}$J/K; these contributions are negligible when compared to the contributions of the translational and rotational degrees of freedom ($\approx$20.775 J/K).}. In the particular case of the Guevara cubic and non-cubic equations, our result provides a better approximation to the observed data, when compared to the virial expansions up to the third order, which was the best prediction to our knowledge. This illustrates that the contribution of molecular interactions beyond the two- and three-particle ones can be relevant under certain circumstances.

This work can be extended in two different directions. One possibility is considering atmospheres that are composed of several gases. Taking into account all possible interactions within a mixture of gases is nontrivial, and the contribution of the latter to the DALR is unknown. The other possibility for future work consists of including intermolecular interactions and vibrations in other approaches to the atmospheric lapse rate, like the so-called ``moist" lapse rate. The role of the latter in the relation between $\eta$ and $\sigma^2$ requires further analysis .

\begin{acknowledgments}

Bogar D\'iaz acknowledges support from the CONEX-Plus programme funded by the Universidad Carlos III de Madrid and the European Union's Horizon 2020 research and innovation programme under the Marie Sk{\l}odowska-Curie Grant agreement No. 801538. J.E.R. acknowledges financial support from the Consejo de Ciencia y Tecnología del Estado de Puebla. This work has been supported by the Spanish Ministerio de Ciencia Innovaci\'on y Uni\-ver\-si\-da\-des-Agencia Estatal de Investigaci\'on under Grant No. PID2020-116567GB-C22.
\end{acknowledgments}

\bibliography{main}

\end{document}